\documentstyle{caosp}

\begin{document}

\pubyear{1998}
\volume{27}

\firstpage{335}
\hauthor{M. S. Dimitrijevi\'c and D. Tankosi\'c}

\htitle{ Stark broadening regularities: Shifts}

\title{Stark broadening parameter regularities and interpolation and
critical evaluation of data for CP star atmospheres research: Stark
line shifts}

\author{Milan S. Dimitrijevi\'c \and Dragana Tankosi\'c}

\institute{Astronomical Observatory, Volgina 7, 11000 Belgrade,
Serbia, Yugoslavia
E-mail: mdimitrijevic@aob.aob.bg.ac.yu}

\date{November 17, 1997}

\maketitle

\begin{abstract} In order to find out if regularities and systematic trends
found to be apparent among experimental Stark line shifts allow the
accurate interpolation of new data and critical evaluation of
experimental results, the exceptions to the established regularities
are analysed on the basis of critical reviews of experimental
data, and reasons for such exceptions are discussed.

We found that such
exceptions are mostly due to the situations when: (i) the energy gap
between atomic
energy levels within a supermultiplet is equal or comparable to the
energy gap to the nearest perturbing levels; (ii) the most important
perturbing
level is embedded between the energy levels of the supermultiplet;
(iii) the forbidden transitions have influence on Stark line shifts.

\keywords{line profile -- atomic data}

\end{abstract}

\section{INTRODUCTION}

Wiese and Konjevi\'c (1982) established that for experimental
Stark widths of non-hydrogenic lines, there are similarities (see
as well references in Wiese \& Konjevi\'c 1982 and Dimitrijevi\'c 1982) of
line widths within a multiplet, a supermultiplet and a transition
array, as well as for analogous transitions of homologous atoms and
ions. They found as
well a systematic behaviour of Stark line widths along spectral series.
The exceptions to these similarities and systematic trends have been
analyzed by Dimitrijevi\'c (1982), who found
that the reasons for such exceptions may be divided in two categories:
(i) irregular atomic energy level structure and (ii) inadequacy of the
model used for the emitter structure. He emphasized as well, that the
simple analysis of Grotrian diagrams for corresponding radiator energy
levels, may be useful for prediction of mutual relations among Stark
widths within multiplets, supermultiplets and transition arrays.
Extending their work of 1982 on Stark widths, Wiese \& Konjevi\'c (1992)
carried out the same kind of research on experimental Stark line
shifts, and showed numerous examples where the same regularities
and systematic trends hold. Similarly as in
Dimitrijevi\'c (1982) for widths, we want to analyze here the
exceptions to the established regularities and systematic trends for
Stark line shifts.

\section{RESULTS AND DISCUSSION}

The exceptions to the established regularities
have been analysed on the basis of critical reviews of experimental
data (Konjevi\'c \& Roberts 1976; Konjevi\'c \& Wiese 1976;
Konjevi\'c et al. 1984ab; Konjevi\'c \& Wiese 1990).
The complete analysis will be published elsewhere. We found that such
exceptions are mostly due to the situations when: (i) the energy gap
between atomic
energy levels within a supermultiplet is equal or comparable to the
energy gap to the nearest perturbing levels; (ii) the most important
perturbing
level is embedded between the energy levels of the supermultiplet;
(iii) the forbidden transitions have influence on Stark line shifts.

The example of Stark line shifts from F I 3s - 3p
(quartets) supermultiplet, illustrates the case when the energy gap
between upper atomic energy levels for particular members of a
supermultiplet is not negligible in comparison to the energy gap to the
most important perturbing
levels. For the 3p$^4$S$^o$ energy level for instance, the influence of the 
upper perturbing levels 4s and 3d, is larger in comparison with this influence
for the 3p$^4$P$^o$
energy level, and the contribution of the 3s energy level is smaller. The effect
of such an energy structure is larger on the shift than on the width,
since all partial contributions to the width are positive while the
contribution of the level 3s as a perturbing level of 4p to the shift
is negative. Consequently, the shift of lines within the 3s$^4$P -
3p$^4$S$^o$ multiplet is larger than the shifts within the 3s$^4$P -
3p$^4$P$^o$ multiplet.

{}

\begin{thebibliography}{}

\bibitem{}{Dimitrijevi\'c, M. S.: 1982, Astron. Astrophys. {\bf 112}, 251.}

\bibitem{}{Djurovi\'c, S., Konjevi\'c, N.: 1988, Z.Phys. D {\bf 10}, 425.}

\bibitem{}{Konjevi\'c, N., Roberts, J. R.: 1976, J. Phys. Chem. Ref. Data
{\bf 5}, 209.}

\bibitem{}{Konjevi\'c, N., Wiese, W. L.: 1976, J. Phys. Chem. Ref. Data
{\bf 5}, 259.}

\bibitem{}{Konjevi\'c, N., Dimitrijevi\'c, M. S., Wiese, W. L.: 1984a, J.
Phys. Chem. Ref. Data {\bf 13}, 619.}

\bibitem{}{Konjevi\'c, N., Dimitrijevi\'c, M. S., Wiese, W. L.: 1984b, J.
Phys. Chem. Ref. Data {\bf 13}, 649.}

\bibitem{}{Konjevi\'c, N., Wiese, W. L.: 1990, J. Phys. Chem. Ref. Data
{\bf 19}, 1307.  }

\bibitem{}{Wiese, W. L., Konjevi\'c, N.: 1982, JQSRT
{\bf 28}, 301.}

\bibitem{}{Wiee, W. L., Konjevi\'c, N.: 1992, JQSRT
{\bf 47}, 185.  }

\end{thebibliography}
\end{document}